\documentclass[12pt]{article}
\usepackage{graphics}

\input epsf
\newcommand{\be}{\begin{equation}}
\newcommand{\ee}{\end{equation}}

\newcommand{\bea}{\begin{eqnarray}}
\newcommand{\eea}{\end{eqnarray}}

\newcommand{\N}{{\cal{N}}}

\begin{document}

\begin{titlepage}

\bigskip
\hfill\vbox{\baselineskip12pt
\vbox{\hbox{UCLA/03/TEP/35}}
\vbox{\hbox{DCPT-03/65}}
\vbox{\hbox{hep-th/0401173}}}
\bigskip\bigskip

\vskip.5cm
\begin{center}
{\Large{\bf Towards the solution to the giant graviton puzzle}}
\vskip1cm
%{alternative title:}\\
%\vskip1cm
%{\Large{\bf Giant gravitons - the good, the bad and the naked}}
\end{center}
%\vskip1cm

\bigskip
\bigskip
\centerline{Iosif Bena$^1$ and Douglas Smith$^2$}
\bigskip
\medskip
\centerline{$^1$\it Department of Physics and Astronomy}
\centerline{\it University of California, Los Angeles, CA\ \ 90095}
\centerline{ iosif@physics.ucla.edu}
\bigskip
\centerline{$^2$\it Department of Mathematical Sciences}
\centerline{\it University of Durham, South Road, Durham, DH1 3LE, UK. }
\centerline{ Douglas.Smith@durham.ac.uk}
\bigskip
\bigskip

\begin{abstract}
In this note we present several ideas toward the solution to the giant graviton 
puzzle -- the apparent multiplicity of supergravity states dual to field theory
chiral primary operators. We use the fact that, for certain ranges of the angular 
momentum, giant gravitons 
can  be mapped into vacua of a dual theory to argue that the sphere and $AdS$ giant 
gravitons have very different boundary descriptions, and that an unpolarized KK 
graviton is unphysical in the regime where giant gravitons exist. We also show that a generic 
boundary state can correspond to different 
giant graviton configurations, which have non-overlapping ranges of validity.

\end{abstract}

\end{titlepage}

\newpage

\section{ Review of the puzzle}

In the $AdS_5 \times S^5$ supergravity dual of $\mathcal{N}=4$ SYM we have three 
distinct types of configurations which correspond to 
a boundary state with large $\mathcal{R}$-charge. One is a graviton circling
the ``equator'' of the $S^5$, the second is a graviton polarized \cite{Myers} into D3 branes 
wrapping an $S^3 \subset S^5$ (the giant graviton or sphere giant graviton) \cite{Susskind}, 
and the third is a graviton polarized 
into D3 branes wrapping an $S^3 \subset AdS_5$ (the dual giant 
graviton or $AdS$ giant graviton.) \cite{SUSYGoliath,aki,dual}

The field theory interpretation of these states is an interesting issue. The
original puzzle was that single-trace chiral primary operators in the field
theory with $\mathcal{R}$-charge $L$ should be dual to single particle states with
angular
momentum $L$ on the $S^5$. The natural candidate for these states is the graviton.
However, at
finite $N$, there is a cut-off in the field theory since there are no
independent single-trace operators with $L > N$, yet there is no obvious upper
bound on the angular momentum of the graviton. The predicted upper bound was
thought to be due to stringy effects and dubbed the ``stringy exclusion
principle'' \cite{Susskind}. Instead, the resolution turned out to be an IR
effect where the gravity dual was identified as the giant graviton. The size of
the $S^3$ which the brane wraps grows with the angular momentum until the upper
bound of $L=N$ is reached where the radius of the $S^3$ reaches the radius of
the $S^5$.

Unfortunately there are two problems with the above picture. The first
is essentially a technical point -- for $L$ of order $N^{2/3}$
the single-trace
operators are no longer orthogonal (even at large $N$). However, this does not affect
the argument since the correct operators are sub-determinant operators \cite{BBNS}
which are also cut off at $L=N$. The second point, which we address in this
paper, is that there is not only the question of whether the extended giant
gravitons should be preferred over the point-like gravitons but that the
extended $AdS$ giant gravitons also carry the same quantum numbers, appear to
have similar properties to the giant gravitons, but crucially have no upper
limit on $L$ since the sphere they wrap can be arbitrarily large within the
$AdS$ spacetime. So, clearly the giant graviton is the one which should 
correspond to the field 
theory state, but how do we rule out the other two states?

The presence of these two extra states has long been a puzzle, and
different 
arguments have been made about their fate. One possible explanation
is that they correspond to two short multiplets which combine to 
form a long multiplet, whose dimension is no longer protected \cite{SUSYGoliath}. 
However, as we will see, this is not what seems to happen.

The existence of many bulk states with polarized branes dual to only one boundary 
state is highly
reminiscent of a similar problem in gauge--gravity dualities. When one 
discusses the supergravity
dual of the $N=1^*$ theory \cite{ps}, one also encounters three bulk vacua dual to one 
gauge theory vacuum. As we will review in the first chapter, a generic field theory 
vacuum can naively have three supergravity duals. One candidate dual bulk contains 
D3 branes polarized into NS5 branes, another one contains D3 branes polarized into 
D5 branes, and the third one has a singularity and does not contain any polarized branes.

Fortunately the solution to this puzzle is known \cite{ps}. The first 
piece of the solution
is that the two candidate duals which contain polarized branes 
have non-overlapping ranges of validity.  The
second piece of the solution is that the vacuum where the D3 branes 
are not polarized is unphysical. Indeed, in \cite{ps} {\em all} the $\N=1^*$ vacua 
(found in the field theory analysis of \cite{vw, dw})
were mapped to supergravity brane configurations, and there is no 
field theory vacuum which is dual to the bulk vacuum with no polarization. 
Therefore, that vacuum has too big an action compared to the polarized configurations,
 and does not contribute to the AdS-CFT duality \cite{witten-98}.

The purpose of this note is to extend this analysis to giant gravitons and 
to show that the
giant graviton puzzle is solved in an essentially identical way. To do this one needs to
make a conceptual jump, from regarding the giant gravitons as {\em states} in the 
4-dimensional boundary theory to regarding them as {\em vacua} of an auxiliary 
theory, which lives on $L$
coincident gravitons in this background. This theory has been discussed in 
\cite{lozano1} and was successfully used in \cite{lozano} to give a microscopic 
description to some of the giant gravitons.\footnote{In this paper we analyze this 
auxiliary theory only implicitly, by relating it via dualities to much 
better understood theories.} When the giant gravitons sit in the ``near-graviton''  
region, they are indeed dual (by the BDHM extension of the AdS-CFT 
correspondence \cite{BDHM}) to vacua of this theory.

Introducing this auxiliary theory into the puzzle makes the dictionary 
between the giant gravitons and the CFT chiral primaries two-step. One first relates the 
states of the CFT to the vacua of the auxiliary theory, and then relates these 
vacua to the giant gravitons. Fortunately, each step is conceptually rather simple. 
The dictionary between the gauge theory states and the vacua of the auxiliary 
theory is one to one, and not hard to guess. As we will see in section \ref{121}, 
this dictionary relates vacua and states of two theories which are 
both strongly coupled when supergravity is weakly coupled, and vice versa. Thus the
fact that it is one to one is quite natural. 

The 3 to 1 degeneracy we 
had in the original dictionary is now mapped into a 3 to 1 degeneracy in the 
map between a vacuum of the 
auxiliary theory and its three candidate dual giant graviton configurations. 
Fortunately, this problem is almost identical to the one solved in \cite{ps}. 

The physics behind the two maps is conceptually rather simple. Like in the 
$N=1^*$ case, the unpolarized configuration (the Kaluza Klein graviton) 
is unphysical, and has no field theory dual 
\footnote{This was also argued in the original giant graviton paper \cite{Susskind}, 
by estimating the action of the KK graviton, and finding it divergent.}.
Moreover, a given chiral primary can have two giant graviton duals, which 
have non-overlapping ranges of validity. The sphere and AdS giant gravitons have 
therefore completely different origins. Intuitively, the sphere giant comes from one
``single particle'' state with angular momentum/$\mathcal{R}$-charge $L$ 
(roughly speaking a single-trace operator of $L$ $\Phi$'s), 
while the AdS giant comes from $L$ ``single particle'' states with angular 
momentum/$\mathcal{R}$-charge one (an operator with a product of $L$ traces of $\Phi$.). 

We should also note that the map we propose matches very well with, and extends 
the proposal of Corley, Jevicki and Ramgoolam (CJR) \cite{CJR} 
for the chiral primaries dual to the AdS and sphere giant gravitons. Although this 
proposal gives a better understanding of the mapping between extended
objects and field theory operators, it does not solve the giant graviton problem, 
since an operator represented by a rectangular Young tableau of size $L \times H$ 
can correspond to either $L$ 
sphere giants of angular momentum $H$, or to $H$ $AdS$ giants of angular momentum $L$. 
As we will see, our proposal maps this degeneracy to the one encountered in $N=1^*$ 
theory, and thus resolves this puzzle rather nicely.

%\footnote{In fact, we first
%constructed the dictionary, based on the intuition form \cite{ps}, and only then became
%aware of the conjectured description of the AdS giant. 

In section~\ref{Tduality} we dualize the 
geometries sourced by KK gravitons in the $AdS \times S$
spacetimes to geometries where the bulk--boundary duality
is very well understood. We then argue that the solution 
corresponding to the unpolarized graviton is nonphysical, 
by using the dual field theory. The point-like graviton with 
angular momentum of order $N$ can also be directly ruled out as a sensible classical
solution \cite{Susskind} since it receives large quantum corrections because of its 
very large energy density. Our T-duality arguments extend this 
to all pure graviton states in the regime where giant 
gravitons exist.

In section~\ref{121} we review the $N=1^*$ duality, (which is the prototypical duality 
between field theories and bulks with many polarized branes), as well as its trivial 
extension to the theory on a large number of D0 branes. We also
present the one to one map between CFT chiral primaries and vacua of the auxiliary 
theory, and thus complete our proposal. In the Appendix we explore the 
ranges of validity of our construction.

Before proceeding we should remark that in the case of the $AdS_5 \times S^5$ 
giant gravitons, the brane configurations
corresponding to the two gravitons are very similar. Therefore, we will initially 
concentrate on the  $AdS_4 \times S^7$ and
$AdS_7 \times S^4$ cases, where the distinction between sphere and $AdS$ 
giant gravitons is more clear, and then argue that the same picture 
extends to the $AdS_5 \times S^5$ case\footnote{The $AdS_3 \times S^3$ case appears to be very different from the other cases \cite{ads3}.}.

\section{Dualizing the giant gravitons}
\label{Tduality}

Perhaps the easiest way to understand the three (giant) graviton states of
angular momentum $L$ is to do
a T-duality along the momentum of the graviton. The resulting static configuration consists 
locally of $L$ F1 strings, in some transverse fields. These F1 strings can 
appear in three incarnations --  by themselves, and polarized into a D4 brane of 
geometry $R^{1,1} \times S^3$ where the $S^3$ can sit in either group of 4 
(picked-out by the background flux) of the 8 transverse directions.

Note that even though the geometry resulting after the T-duality is singular 
at the poles of the sphere, we know that string theory on that background makes sense. 
The singularity of the T-dual supergravity solution comes from the fact that 
winding modes near the origin can 
shrink to zero size, and thus it is an artifact, signaling the breakdown of
the supergravity approximation to string theory. However, as we will see in the 
Appendix, the auxiliary theory describes the giant gravitons only in a region near 
the equator. Thus, the breakdown of the
supergravity approximation happens in a region away from the F1 strings we are
analyzing.

Near the F1 strings, one cannot consider them any more as a perturbation on
the geometry. They become a {\em source} for the geometry. The geometry near the
strings is the near-horizon F1-string geometry perturbed 
with some transverse fluxes (coming from the 5-form field strength of the
$AdS_5 \times S^5$ geometry), which can cause the strings to be polarized 
into D4 branes. The M-theory lift of this geometry is very reminiscent of the
geometry dual to the massive flow of the M2 brane worldvolume theory \cite{m2}. 
In fact, it
is quite easy to see (using the fact that both geometries allow brane polarization and that
both are supersymmetric) that this M-theory lift is the
massive $AdS_4 \times S^7$ flow geometry in which M2 branes in 
transverse fields polarize into M5 branes.

One can also examine the M-theory giant gravitons, and see that by 
dimensionally reducing them along their momentum, they correspond locally
to D0 branes polarized into D2 branes and NS5 branes respectively. This geometry is also
the gravity dual of the field theory on the D0 branes perturbed by a chiral multiplet
mass. Like in the previous case, the IIA geometry obtained by reducing the M-theory
giant gravitons along their momentum is singular at the poles. However, this only
signifies the breakdown of the IIA supergravity approximation, and the background
makes perfect sense if one considers the full M-theory. As before, the physics we are 
interested in happens in a region away from where IIA supergravity breaks down.

The theory on the D0 branes is just supersymmetric matrix quantum mechanics. The 
background fluxes in which these D0 branes sit are a transverse NSNS 3-form 
field strength and a transverse RR 6-form field strength. It is not hard to 
see\footnote{Either by T-dualizing 3 times the 
Polchinski-Strassler setup, or by analyzing the Non-Abelian Born Infeld action 
of the D0 branes} that the effect of these fluxes on the D0 branes is to induce a mass 
for 3 of the chiral multiplets of the supersymmetric matrix quantum mechanics. The 
supergravity dual of the perturbed theory  has now many vacua, which contain 
polarized branes. The structure of these vacua is identical for any Dp branes 
put in a transverse $H_3$ and $F_{6-p}$. The D3 brane case has been analyzed in \cite{ps}
and the D2 brane case has been analyzed in \cite{d2}.

In the next section we review the example of D3 branes in transverse $H_3$ and $F_3$. This 
is the best understood case where one boundary vacuum naively corresponds to three 
bulk configurations. In that case, the dual field theory intuition helps us understand very 
well how this discrepancy is resolved. We then use the fact that the D0 setup is 
related to the D3 setup case by T-duality to argue that this resolution extends to the 
D0 brane case, and consequently to the giant gravitons.

\subsection{The map between giant gravitons and vacua of the D0 brane theory}

As we explained in the introduction, in the string theory dual of 
the $\mathcal{N}=1^*$ theory one generically has vacua 
containing D3 branes polarized into D5 branes and NS5 branes. Brane polarization
happens because the D3 branes are placed in transverse RR and NSNS 3-form fluxes.

Let us first examine the two vacua corresponding to $L$ D3 branes polarized into one
D5 brane and one NS5 brane respectively. The $\mathcal{N}=1^*$ classical F-term constraints 
are $[X^i,X^j] = \epsilon^{ijk} X^k$. The D5 vacuum corresponds to
the  maximally  Higgsed classical solution, where the $X^i$ are the generators of the
$L \times L$ irreducible representation of SU(2). The NS5 vacuum
corresponds to the classical solution $X^i= 0 \times 1 \!\!\!\! 1$. Thus the $X^i$ 
can be thought of as the generators of the product of $L$ trivial representations of SU(2).
Quantum effects make this state acquire a nonzero $<X^2>$, which can be interpreted 
in supergravity as the D3 branes polarizing into an NS5 brane. In \cite{ps} it was shown 
that this polarization is caused by nonperturbative effects -- indeed, the presence 
of NS5 branes in the bulk corresponds  to confinement in the field theory, and the size 
of the NS5 branes gives the field theory mass gap.

As we can see, the two single-brane states come from completely different 
classical states. This phenomenon is generic to all polarizations of D
branes. The polarization pattern is Dp $\rightarrow $D(p+2) and
Dp $\rightarrow $ NS5. The 
Dp $\rightarrow $D(p+2) configuration corresponds in the classical limit 
to the $L \times L$ irreducible representation of SU(2). The 
Dp $ \rightarrow  NS5$ configuration is made of $L$ classical objects  (corresponding 
to a product of $N$ trivial representations) which via nonperturbative quantum effects 
create one NS5 brane.
\begin{center}
\begin{figure}[h]
\centerline{\scalebox{.75}{\includegraphics{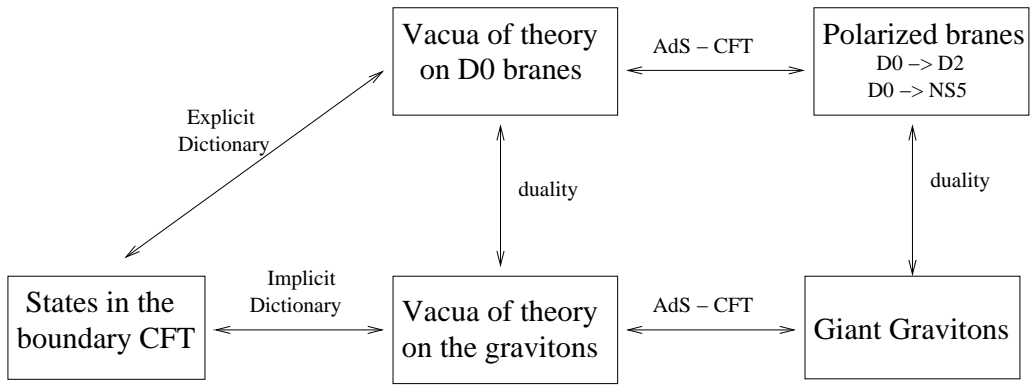}}}
\caption{The dualities behind of the proposed solution.}
\label{DualityDiagram}
\end{figure}
\end{center}
\vspace{-1cm}
 
Once we have established this correspondence, we can go ahead and analyze a more 
general classical configuration, and see what it corresponds to in the supergravity dual.
The various duality relations are sketched in figure~\ref{DualityDiagram}.

For example, a classical configuration in which the $X^i$ are the generators of the 
product of $L/k$ SU(2) irreducible representations of dimension $k$ corresponds in the bulk 
to polarization into $k$ D5 branes. However, this configuration could also correspond to 
polarization into $L/k$ NS5 branes. What solves this apparent puzzle is the fact that 
the two configurations have non-overlapping ranges of validity \cite{ps}. Thus, 
for $k^2 \gg {L \over g} $ the dual bulk configuration has NS5 branes, while for 
$k^2 \ll {L \over g} $ the dual bulk configuration has D5 branes \footnote{This is explained in \cite{ps}, eqns. (83-85).}. This gives a one to 
one map between classical gauge theory configurations 
and bulk configurations with D5 and NS5 branes. 

We should also note that there is 
no field theory vacuum corresponding to the geometry without polarized branes. 
It is quite likely that this configuration has a naked singularity, and thus it 
has too large an action to contribute in the AdS-CFT 
correspondence. The fact that the bulk and boundary analysis of the vacua of 
the $\N=1^*$ theory match so precisely, and that there is no boundary vacuum dual 
to this geometry is a very strong argument that this state indeed does not 
appear in the AdS-CFT correspondence. This shows that KK graviton is probably 
singular when a giant graviton
with the same angular momentum exists. When this 
giant graviton is smaller than the string scale, the KK graviton is physical.

The $N=1^*$ classical analysis of the vacua extends trivially to the D0 brane 
case\footnote{The only difference is that for D0 branes there are no oblique states, 
and therefore vacua do not proliferate as one goes from classical to quantum.}. 
Thus, the maximally Higgsed irreducible representation corresponds to the D0 branes becoming 
one D2 brane, and the product of $L$ trivial representations corresponds to the D0 
branes polarizing into one NS5 brane. These two configurations are the reduction 
of the biggest M-theory sphere and $AdS$ giant gravitons. A product 
representation 
can again be interpreted as the dual of $k$ giants or $L/k$ $AdS$ giants, depending 
on $k$ and the coupling constant.

Thus, we have a very clear one to one map between 
giant graviton configurations and vacua of the auxiliary gauge theory on the 
D0 branes. 
The map is naively three to one, but we have seen that the configuration with 
no polarization (dual to the KK graviton) is excluded, and the two configurations 
dual to the same gauge theory vacuum have non-overlapping ranges of validity.

\section{The one to one map}
\label{121}

% Having established the map between giant graviton configurations and vacua of 
% the auxiliary gauge theory on the D0 branes one can 
% also present a map between 
% vacua of this theory and states on the boundary theory. The essential physical 
% intuition one uses for this map comes from the picture presented in the 
% previous chapter. 

The auxiliary theory description presented in the previous section 
allows one to think about the D0
$\rightarrow$ NS5 state as corresponding classically to $L$ particles of angular
momentum one, which form a bound state because of quantum effects. In a similar way, 
the D0 $\rightarrow$ D2 state would correspond 
classically to one state of angular momentum $L$. 

This picture is furthermore 
supported by the fact that the state with $L/k$ $AdS$ giants and the state 
with $k$ sphere giants correspond to the same auxiliary theory configuration. 
Therefore, $k$ giant gravitons of one kind correspond to $k$ states each containing $L/k$ 
particles, while $k$ giant gravitons of the other kind correspond to $L/k$ 
states each containing $k$ particles. Hence, only one of the maximal giant 
gravitons found in supergravity corresponds to a
gauge theory ``single particle'' state, while the other
describes a ``multi-particle'' state. 

To summarize, the main lesson we can draw from the auxiliary theory description of the
M-theory giant gravitons, is that $k$ sphere giants of angular 
momentum $L/k$ correspond to the same boundary configuration as $L/k$ dual 
giants of angular momentum $k$, but in a different
regime of the parameter space. This indicates that the maximal sphere and maximal 
$AdS$ giants correspond to classical configurations that roughly speaking can be thought of as 
``single particle'' and ``multi-particle'' states. 
We would like now to use this intuition to discuss the $AdS_5 \times S^5$ giant gravitons. 

As we show in the Appendix, 
the auxiliary map of section \ref{Tduality} does not 
extend to the case when the $\mathcal{R}$-charge is comparable to $N$, basically 
because the giant gravitons become bigger than the near-horizon region they 
source, and are thus not described any more by the auxiliary theory. 
However, we do not believe this affects the picture above. It would be rather 
strange if the fact that two different giant gravitons correspond 
to the same boundary state changed, especially in the 
$AdS_5 \times S^5$ case were one can go out of the ``near graviton'' region (\ref{gs}) by 
changing a continuous small parameter - $g_s$. 

Through the usual correspondence, the
number of particles becomes a number of
traces in the field theory operator, at least for $\mathcal{R}$-charge 
small\footnote{When the $\mathcal{R}$-charge becomes comparable to $N^{2/3}$, 
this picture starts getting corrected \cite{BBNS} -- the single traces 
become sub-determinants.}  
compared to $N^{2/3}$. Thus, for small $L$, a maximal sphere giant should correspond 
to a single trace operator, while the maximal $AdS$ giant should correspond to a 
chiral primary operator with $L$ traces. Moreover, given a total angular momentum 
$L$, $k$ sphere giants and $L/k$ $AdS$ giants correspond to the same CFT state.

We can see that the picture which emerges from our description 
matches very well with the proposal of 
Corley, Jevicki and Ramgoolam for the chiral primaries dual to the 
$AdS$ and sphere giant gravitons \cite{CJR}. According to this proposal, 
the sphere giant gravitons
and $AdS$ giant gravitons are both dual to $\N=4$ SYM 
chiral primaries of  $\mathcal{R}$-charge $L$. An efficient method to index 
the chiral primaries of this field theory is to associate to each 
primary a $U(N)$ Young Tableau.
For $L \ll N$ the number of traces in each operator corresponds to the number 
of columns of the Young tableau, and the number of fields in each trace 
corresponds to the number of boxes in the corresponding column. For larger $L$ mixing 
becomes important, and the dictionary becomes more involved \cite{BBNS}. The
total number of boxes in the Young tableau gives the $\mathcal{R}$-charge
$L$ of the operator, which maps to the total angular momentum of the dual
supergravity state.

Now the proposal in \cite{CJR} was that a single column of $L$ boxes, i.e.\ the
totally anti-symmetric rank $L$ representation, corresponds to the operator dual to
a single sphere giant graviton with angular momentum $L$.
Clearly these states fit in with the ``stringy exclusion principle''
as they are both cut off at $L=N$. Several columns of equal length would
correspond to several giant gravitons (or a single multiply wrapped giant
graviton.) Similarly, a single row of $L$ boxes, i.e.\ the totally symmetric
rank $L$ representation, was proposed to be dual to a single $AdS$ giant
graviton with angular momentum $L$. For small $L$ the corresponding field 
theory operator contains a product of $L$ single traces\footnote{The gauge group is
taken to be U($N$) in \cite{CJR} rather than SU($N$). However, on the
supergravity side, this difference may not be apparent without considering
quantum corrections.}. This confirms the heuristic picture of the dual giant 
as a bound state of $L$ particles with angular momentum $1$, which emerged 
from the auxiliary theory

By combining the auxiliary theory description of giant gravitons 
with the CJR proposal, one can formulate a map between field theory operators and auxiliary 
theory vacua, which naturally extends this proposal. Thus, a CFT operator described by a 
Young tableau with $k$ columns of lengths $L_i$ corresponds to 
the same giant graviton configuration as a 
classical vacuum of the auxiliary theory in which the scalars are in the 
SU(2) representation which is the
product of $k$ irreducible representations of size $L_i \times L_i$.
The number of boxes of the Young tableau is the total angular momentum 
of the giant graviton, which in the auxiliary theory gives the rank of the 
gauge group, and is equal to the sum of the $L_i$'s. The 
completely vertical Young tableau is linked by this map to the maximally Higgsed 
vacuum, which
corresponds to the $D0-D2$ polarization channel. The completely horizontal 
Young tableau 
is linked to the product of $L$ trivial representations, which corresponds to the 
$D0-NS5$ polarization. 

The CJR proposal, although giving a better understanding of the mapping between giant 
gravitons and chiral primaries, still does not solve the giant graviton problem, 
since an operator represented by a rectangular Young tableau of size $L \times H$ 
can correspond to either $L$ sphere giants of angular momentum $H$, or 
to $H$ $AdS$ giants of angular momentum $L$. 
Our proposed map solves this problem very easily, by mapping any Young tableau 
to a classical vacuum of the gauge theory on 
the D0 branes, and using the fact that the two supergravity duals of this vacuum 
have non-overlapping ranges of validity (at least
for rectangular Young tableaux -- and we propose this is true in general.) So for a given range of 
parameters there is always only one valid supergravity solution per chiral primary.

Although our map agrees with the CJR proposal for the 
maximal sphere and $AdS$ giant, it seems to differ a bit  for less symmetric cases. 
For example, in \cite{CJR} a Young tableau with two columns of lengths $L_1$ and 
$L_2$, such that $0 < L_2-L_1 \ll L_2$ was argued to correspond to two giant gravitons of 
momentum $L_1$, and to one KK graviton of momentum $L_2-L_1$. According to our map, 
this Young tableau should correspond to two  giant gravitons of momenta $L_1$ and 
$L_2$, and no free KK gravitons. As we have explained, a KK graviton of 
momentum $L_2-L_1$ should be 
unphysical when the giant graviton of the same momentum is a valid solution. 
If $L_2-L_1$ is very small, and the corresponding giant graviton does not exist, the
KK graviton should correspond 
to a Young tableau column of length $L_2-L_1$ . It would be interesting to see 
if this could be independently checked.

We should also remark that the duals of the sphere giant gravitons are different 
in the case of $AdS_7 \times S^4$ and $AdS_4 \times S^7$. In one case the sphere 
giant is dual to a D2 brane, and in the other to an NS5 brane. Our arguments imply 
that one of the giant gravitons corresponds to the single particle state and the 
other one to the many particle state. However, we cannot say which of the 2 giant 
gravitons is the ``single trace'' one, essentially because the dualities used to 
get to the auxiliary theory description are strong-weak dualities, which can cause
large mixing between single trace and multiple trace operators \cite{go}
(essentially in the same way in which 
S-duality in the $N=1^*$ theory interchanges the D5 and the NS5 vacua). 
The same difficulty persists to the $AdS_5 \times S^5$ case, where the $AdS$ and 
sphere giant gravitons have the same brane content. One can also see that it is 
difficult to distinguish which of the two giant gravitons is the dual of a single 
particle state because S-duality maps each brane to itself. 

What seems however to be very generic is that when we have two polarization 
channels, only one channel corresponds to a single particle state/single 
trace operator, while the other one corresponds to a many particle state/multi 
trace operator. Thus the sphere giant and AdS giant have completely different 
field theory duals, despite their similarity in supergravity. 

As we have explained in the previous section, the M-theory lift of the T-dual of the IIB
near-graviton geometry  is the massive flow of the M2 brane theory, which contains M2 
branes polarized into M5 branes. The sphere and $AdS_5$ giants correspond to the 
two orientations of the polarization planes. 

According to our dictionary, 
the two duals of a gauge theory Young Tableau of size $k \times {L\over k}$ would be 
a state with $k$ M5 branes of M2 charge $L/k$ each, and a state with $L/k$ M5 branes of 
M2 charge $k$ each. However, it was shown in \cite{m2} that a state with a total of $L$ 
M2 branes polarized into $k$ M5 branes only has a valid description for 
$$ k^2 < L. $$
Thus, the two  duals of the gauge theory state described above  have 
non-overlapping ranges of validity. 
We should note that this is a rather nontrivial check for our conjectured dictionary, 
given that the bound above comes from the M5 brane action, which does not have 
many things in common with chiral primaries of 4-dimensional $\N=4$ Super Yang Mills.

Another prediction of our map is that there is an upper bound on the total number 
of $AdS$ giants. Indeed, the number of $AdS$ giants and the angular momentum of the sphere 
giants have the same auxiliary theory interpretation, and thus both should be cut off at $N$. 
To see that this is the case we recall that the $AdS$ giant is a spherical brane domain wall 
in $AdS$, and therefore the flux which supports the $AdS \times S$ geometry jumps across the brane. 
Let us now imagine having exactly $N$ $AdS$ giant gravitons. The flux inside this configuration 
is zero, and so if one tries to form another $AdS$ giant graviton there is no flux to support it from collapsing. If there are more than $N$ $AdS$ giants, the flux in the region inside them changes sign, and pulls down some of the $AdS$ giants, reducing the number below $N$.

\section{Conclusions}

We have examined the expansion of point-like gravitons into giant gravitons, and 
have argued that when giant gravitons exist, the 
point-like gravitons have no field theory dual. Most probably they are not valid solutions of 
supergravity. Moreover, we have argued  
by analogy to the study of the $\N=1^*$ theory by Polchinski and Strassler \cite{ps}, 
that the two different types of giant graviton have distinct interpretations in the field
theory. 

In the supergravity description the giant graviton and dual giant
graviton appear to be similar objects, arising from the coupling of the
point-like graviton to a background field, either electrically or magnetically.
However, we have presented arguments that only one of these two expanded 
configurations corresponds classically to a single particle/single trace
state, while the other one corresponds classically to many particles, which form a bound 
state via quantum effects. This interpretation also ties in nicely with the map between giant gravitons and field theory operators via Young tableaux \cite{CJR}.

Moreover, we have shown that a collection of 
$L/k$ $AdS$ giants with angular momentum $k$ and a collection of $k$ sphere giants with 
angular momentum $L/k$ correspond to the auxiliary theory vacuum, and hence to the 
same CFT chiral primary, but in different 
regimes of parameters. We have thus presented a solution to the
problem that there are apparently many more configurations involving giant and
dual giant gravitons than there are appropriate dual field theory operators.

$ $\\
{\bf Acknowledgements}: We would like to thank Allan Adams, Ofer Aharony, 
Dan Kabat, Per Kraus, Don Marolf, Joe Polchinski, Radu Roiban and 
Simon Ross for stimulating discussions. We also thank the Aspen Center for Physics for 
support during the initial stages of this project. The work of IB was 
supported in part by the NSF grant  PHY00-99590.

\newpage

\appendix

\section{Ranges of validity}
\label{ranges}

In this appendix we explore the range of angular momenta when the giant graviton 
states can be described as vacua of the auxiliary theory living on the gravitons. 
In order for this to happen, the size $r_g$ of a giant graviton of momentum $L$ 
must be smaller than the size of the near-horizon region of the gravitons $r_0$. 
Moreover, when the size of the giant graviton becomes smaller than the string or 
Planck scale, it makes sense to treat it as a KK graviton. 

To estimate $r_0$ we use the fact that in 10 dimensions, the harmonic 
function sourced by $L$ gravitons is of the form 
\be
Z - 1 \sim {{L \over R} g_s^2 \over r^6} = {r_0^{6} \over r^{6}},
\label{est10}
\ee
while in 11 dimensional supergravity the harmonic function is
\be
Z - 1 \sim {{L \over R} \over r^7} = {r_0^{7} \over r^{7}},
\label{est11}
\ee
The total energy in the gravitons is ${L \over R}$, and the extra factor of $g^2$ 
in (\ref{est10}) comes from using string units instead of Planck units.

For $AdS_7 \times S^4$, size of the near-graviton region is (\ref{est11})  
\be
r_0^7 \sim {L \over R} \sim {L \over N^{1/3}}
\ee
while the size of a giant graviton of angular momentum $L_i$ is \cite{Susskind}
\be
r_g = {L_i R \over N}  \sim {L_i \over N^{2/3}}
\ee
Therefore, a state containing a single giant graviton is described by the auxiliary theory 
for  $L < N^{13/18}$. Moreover, the requirement that the size of the giant graviton 
be bigger than the Plank length gives a lower bound on $L_i$:   $L_i > N^{2/3} $.
If we have more giant gravitons, the near-graviton region increases, while
the size of the giants remains the same, so the range described by the auxiliary 
theory increases.

For $AdS_5 \times S^5$, the size of the near-graviton region is (\ref{est10})  
\be
r_0^7 \sim {L g_s^2 \over R} 
\ee
where $R^4 \sim  g_s N$. The size of the giant graviton of angular momentum $L_i$ 
is \cite{Susskind}
\be
r_g^2 = {L_i R^2 \over N}  
\ee
Therefore, a state containing a single giant graviton is described by the auxiliary theory for  
\be
L^2  < N R/l_s \sim  N^{5/4} g_s^{1/4}.
\label{gs}
\ee
The requirement that the size of the giant graviton be bigger than the string
length gives the lower bound: $L^2 > N g_s^{-1} $.

For $AdS_4 \times S^7$, the horizon size is (\ref{est11})  
\be
r_0^7 \sim {L \over R} \sim {L \over N^{1/6}}
\ee
while the size of a giant graviton is \cite{Susskind}
\be
r_g^4 = {L_i R^4 \over N}  \sim {L_i \over N^{1/3}}
\ee
Therefore, a state with one giant graviton is  described by the auxiliary 
theory for  $L < N^{5/9}$. The lower bound on $L$ is $L > N^{1/3}$.

We should note that the window of parameters 
in which giant gravitons are described by the auxiliary 
theory grows with $N$. Morever, the window can be made larger if one considers states 
with many  giant gravitons. This window 
covers but a fraction of the available parameter space. However, as we have 
explained in section~\ref{121} the basic physics which the auxiliary theory 
analysis reveals remains valid throughout the whole parameter space.

\end{document}